# Real-Space Visualization of Frequency-Dependent Anisotropy of Atomic Vibrations

Xingxu Yan[1]†, Paul M. Zeiger[2]†, Yifeng Huang[3], Haoying Sun[4], Jie Li[3], Chaitanya A. Gadre[3], Hongbin Yang[1], Ri He[5], Toshihiro Aoki[6], Zhicheng Zhong[5], Yuefeng Nie[4], Ruqian Wu[3], Ján Rusz[2]*, Xiaoqing Pan[1,3,6]*

[1]Department of Materials Science and Engineering, University of California, Irvine, CA 92697, USA.

[2]Department of Physics and Astronomy, Uppsala University, P.O. Box 516, Uppsala 75120, Sweden.

[3]Department of Physics and Astronomy, University of California, Irvine, CA 92697, USA.

[4]National Laboratory of Solid State Microstructures, Jiangsu Key Laboratory of Artificial Functional Materials, College of Engineering and Applied Sciences, Nanjing University, Nanjing 210093, China.

[5]Key Laboratory of Magnetic Materials Devices & Zhejiang Province Key Laboratory of Magnetic Materials and Application Technology, Ningbo Institute of Materials Technology and Engineering, Chinese Academy of Sciences, Ningbo 315201, China.

[6]Irvine Materials Research Institute, University of California, Irvine, CA 92697, USA.

*Correspondence to: jan.rusz@physics.uu.se (J.R.); xiaoqing.pan@uci.edu (X.P.).
†These authors contributed equally to this work.

**The underlying dielectric properties of materials, intertwined with intriguing phenomena such as topological polariton modes and anisotropic thermal conductivities, stem from the anisotropy in atomic vibrations[1–6]. Conventionally, X-ray diffraction techniques have been employed to estimate thermal ellipsoids of distinct elements, albeit lacking the desired spatial and energy resolutions[6–8]. Here we introduce a novel approach utilizing the dark-field monochromated electron energy-loss spectroscopy for momentum-selective vibrational spectroscopy, enabling the cartographic delineation of variations of phonon polarization vectors. By applying this technique to centrosymmetric cubic-phase strontium titanate, we successfully discern two types of oxygen atoms exhibiting contrasting vibrational anisotropies below and above 60 meV due to their frequency-linked thermal ellipsoids. This method establishes a new pathway to visualize phonon eigenvectors at specific crystalline sites for diverse elements, thus delving into uncharted realms of dielectric, optical, and thermal property investigations with unprecedented spatial resolutions.**

Collective atomic vibrations in a periodic lattice form quasi-particles which are termed phonons according to the quantum mechanical theory. Phonon anisotropy, the difference of phonon spectra along different crystalline orientations, exists in low-symmetry system and causes the anisotropy of the dielectric function[1], optical responses[2,3], and thermal properties[4] of crystalline materials. For instance, the anisotropic phonon modes in Reststrahlen bands yield anisotropic in-plane polaritons propagating along the surface of two-dimensional (2D) α-$MoO_3$ films[2,9] and topological polaritons occurring in twisted α-$MoO_3$ bilayers[3]. The large-scale 2D $MoS_2$ films with random interlayer rotations produce a ratio of in-plane and out-of-plane thermal conductivities close to 900 owing to the phonon anisotropy of layered structure[4]. Traditionally, angle-resolved polarized Raman spectroscopy is applied to measure the phonon spectra along different crystalline directions[9]. However, optical spectroscopy along with other phonon-detecting methods including inelastic neutron scattering and X-ray scattering can only offer a macroscopic measurement with a poor spatial resolution of a few micrometers[10]. It is conceivable that vibrational anisotropy could occur at the atomic level due to a locally reduced point group symmetry, despite the higher space group symmetry of the entire crystal[5]. Anisotropic atomic vibrations can fundamentally affect the thermal expansion and optical responses of crystals and are typically estimated as thermal ellipsoids using diverse diffraction techniques[6–8,11]. However, these approaches encountered drawbacks of lacking adequate frequency and spatial resolutions and cannot unveil the atomic level anisotropy for individual atoms in crystals. Experimental direct observations of phonon polarization vectors and vibrational anisotropies at the atomic level have yet to be achieved.

Very recently, an emerging monochromated electron energy-loss spectroscopy (EELS) method in a scanning transmission electron microscope (STEM) demonstrates a combination of sub-nanometer spatial resolution and sub-10 meV energy resolution for probing vibrational states and phonon modes of both organic and inorganic materials[12]. This method can detect local phonon modes at crystalline imperfections such as point defects[13,14], stacking faults[15], interfaces[16–20], and grain boundaries[21,22]. It is also possible to map out the intensity fluctuation of entire phonon spectra associated with individual atoms[19,23,24]. However, distinguishing diverse elements, crystalline positions, or atomic vibrational anisotropies has not been realized. Here, we demonstrate a momentum-selective dark-field vibrational EELS method with the capability of distinguishing various elements at different crystalline sites to directly image the atomic feature of phonon modes in narrow energy ranges. In a centrosymmetric strontium titanate ($SrTiO_3$ or STO in short), we unveil the frequency-dependent anisotropy of oxygen vibrations and isotropy of Sr and Ti vibrations in varied crystalline sites previously inaccessible with other experimental approaches.

The cubic phase STO is an incipient ferroelectric material, exhibits unique dielectric[25–27] and optical properties[28,29], and has been widely used for benchmarking new experimental methodologies, including advanced imaging and spectroscopic approaches[30–32]. The phonon structure of STO dictates the antiferrodistortive phase transition[33,34], nanoscale dielectric dead layer[25,26], and even influences the high-temperature superconducting behavior in the grown FeSe layers via the electron-phonon coupling effect[35]. Therefore, it is important to investigate the atomic vibrational spectra and the frequency-dependent thermal ellipsoids to understand the underlying mechanism of above behaviors. Freestanding STO films were used in this study to take advantage of their well-controlled composition and uniform thickness without the substrate-induced clamping effect[36]. We transferred 50-unit-cell (u.c.) thick freestanding STO films (~20 nm) on top of a few-layer graphene film (3-6 L, Extended Data Fig. 1) to mitigate the drifting and charging issues in order to achieve high quality vibrational EELS hyperspectral imaging datasets with atomic resolution (see more details in Methods). We first performed dark-field vibrational EELS experiments using a high energy resolution Nion microscope as illustrated in Fig. 1a to obtain localized phonon signals[13,14,23]. An electron beam with a convergence semi-angle of 33 mrad is formed to raster scan on the STO film with a 1.5 Å spatial resolution[15]. Then, transmitted and scattered electron beams are deflected via projector lenses to shift the central disk away from the EELS entrance aperture (EEA, a collection semi-angle of 25 mrad) by an angle of 62 mrad to set up the dark-field EELS condition. The graphene film produces negligible imaging contrast and phonon signals in the energy range of interest (see more details in Extended Data Fig. 2).

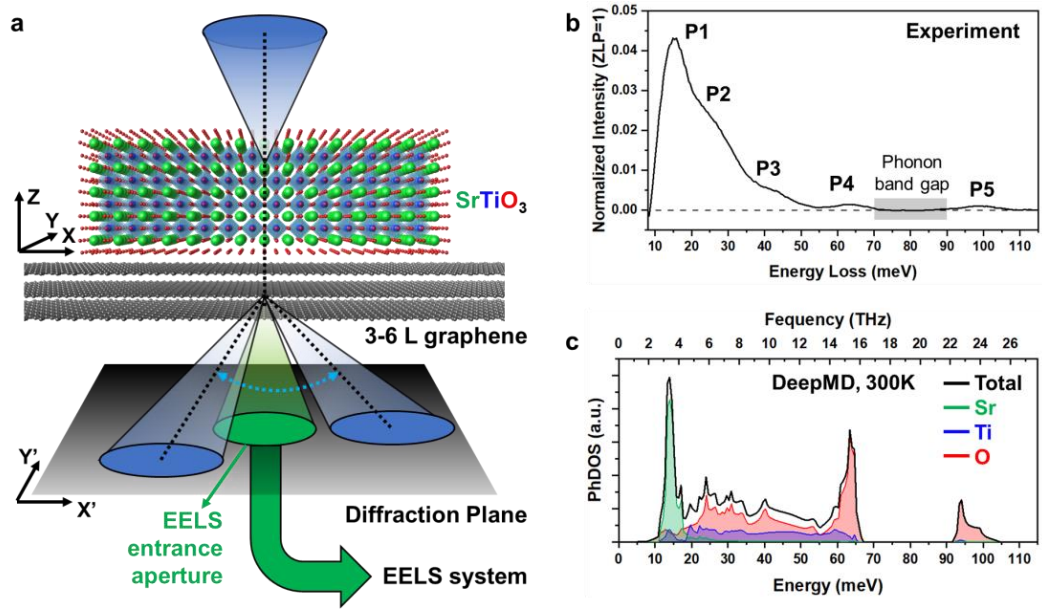

**Fig. 1| Acquisition of vibrational spectrum of free-standing $SrTiO_3$ film. a,** Schematic of a momentum-selective dark-field EELS experimental setup. A convergence semi-angle of 33 mrad is used to form an electron probe. The STO film is attached to a suspended few-layer graphene. Green, blue, red, and gray balls denote Sr, Ti, O, and C atoms respectively. The position of

diffraction pattern in the diffraction plane is shifted via changing the post-specimen projector system to modify the relative position of EELS entrance aperture (EEA, the green circle). The dark-field EELS signal is collected from the EEA and forms vibrational spectra in the EELS system. **b,** The background-subtracted vibrational spectrum of STO averaged from a 2 nm × 2 nm region. **c,** DeepMD-based PhDOS of STO and projected PhDOS of Sr, Ti, and O atoms at 300 K. The phonon band gap between 70–90 meV also appears in the experimental spectrum and is used to conduct reliable background subtractions.

Figure 1b presents a background-subtracted vibrational spectrum of STO (see more details in Methods). There are five major peaks labeled as P1–P5 with peak centers of 15 meV, 27 meV, 43 meV, 63 meV, and 99 meV, respectively. We performed molecular dynamics (MD) simulations with a newly developed deep learning potential[37] to calculate the phonon dispersion curves (Extended Data Fig. 3e) and phonon density of states (PhDOS, Fig. 1c) of cubic phase STO with non-analytical correction at 300 K and verified the reliability of such a method by comparing with those obtained from first-principles calculations (Extended Data Fig. 3). There are four pairs of longitudinal optical (LO) and doubly degenerate transverse optical (TO) modes in the phonon band structure of cubic phase STO[38,39]. Thus, P5 peak is assigned to LO4 mode with predominantly oxygen vibrations, P4, P3, and P2 are assigned to TO4/LO3 modes, TO3/LO2 modes, and TO2/LO1 modes with vibrations of both Ti and O atoms, respectively, and P1 is assigned to the mixture of TO1, longitudinal acoustic (LA), and transverse acoustic (TA) modes with mainly Sr vibrations. It is worth noting that the TO1 mode at 15.8 meV is the so-called soft phonon mode governing the antiferrodistortive phase transition at 105 K[33,34]. Compared to previous studies[18,21], A-site Sr-related phonon modes below 20 meV can be well observed and resolved in our results. Therefore, the experimental vibrational spectrum captures most of the phonon modes in the entire 10–110 meV energy range thanks to the correction of high order EELS aberrations (see more details in Methods). According to the atom projected PhDOS, phonon modes in different energy ranges consist of distinct vibrational states from different elements of Sr, Ti, or O.

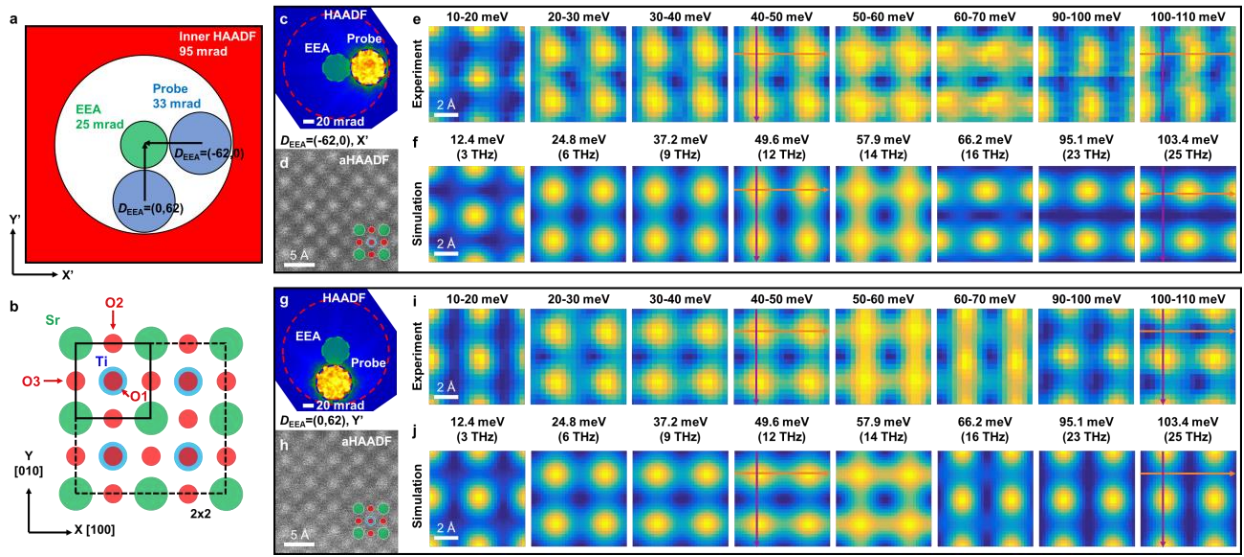

**Fig. 2| Atomic resolution vibrational signal mapping in different energy ranges with two orthogonal displacement directions of EEA. a,** Schematic of diffraction plane showing the movable probe (blue circle), a fixed EEA (green circle), and a HAADF detector (red area). The

horizontally and vertically shifted probes offer $D_{EEA}$ of (-62, 0) and (0, 62), respectively. The unit of $D_{EEA}$ is mrad and is omitted in the main text. **b,** Atomic structure of STO along [001] direction. The boxes with black solid lines and dashed lines denote single u.c. and 2×2 u.c., respectively. The oxygen overlapped with Ti is labeled as O1, while the apical and equatorial oxygens are labeled as O2 (top and bottom) and O3 (left and right), respectively. **c-f,** Experimental diffraction pattern (**c**), aHAADF-STEM image (**d**), experimental vibrational signal mapping averaged from 24 individual u.c. in different energy ranges (from 10-20 meV to 100-110 meV, **e**) and simulated vibrational signal mapping at representative energies (**f**) with $D_{EEA}$ = (-62, 0). **g-j,** Experimental diffraction pattern (**g**), aHAADF-STEM image (**h**), experimental vibrational signal mapping averaged from 30 individual u.c. in different energy ranges (from 10-20 meV to 100-110 meV, **i**) and simulated vibrational signal mapping at representative energies (**j**) with $D_{EEA}$ = (0, 62). The mapping results of 70-80 meV and 80-90 meV are not present here due to the extremely low signals in the phonon band gap. Mapping data are duplicated from averaged single u.c. results to 2×2 u.c. for visual clarity. Simulated mapping results are blurred with a 2D Gaussian of width 2 Å. Scale bars in (**c, g**) are 20 mrad. Scale bars in (**d, h**) are 5 Å. Scale bars in (**e, f, i, j**) are 2 Å.

Then, we performed atomic resolution vibrational EELS mapping on a small region covering a few u.c. (see more details in Methods and Extended Data Fig. 4) in a controllable manner. Rather than shifting the diffraction pattern to an arbitrary direction in previous studies, we moved the central disk along certain crystalline directions of the observed sample with respect to the position of EEA to selectively probe phonon eigenmodes along different polarization directions. The displacement vector $D_{EEA}$ is defined from the probe center to the EEA center to quantify the momentum exchange of scattered electrons as shown in Fig. 2a. [100] and [010] directions of STO are labeled as X and Y axes in the real space (Fig. 2b), and parallel to X' and Y' axes in the reciprocal space (Fig. 2a). We shifted the diffraction pattern along either X' axis with $D_{EEA}$ = (-62,0) in Fig. 2c or Y' axis with $D_{EEA}$ = (0,62) in Fig. 2g. Since the central disk of diffraction pattern under dark-field EELS condition deviates from the center of the HAADF detector, the acquired HAADF-STEM images are so-called asymmetric HAADF (aHAADF) STEM images[13,14,23]. Under both shifts along X' and Y' directions, the aHAADF-STEM images (Figs. 2d and 2h) display the atomic structure of STO with the brightest atoms as Sr and less brighter atoms as Ti, but oxygen atoms are invisible. The background-subtracted spectra of one hyperspectral imaging dataset (Extended Data Fig. 5a) show a strong intensity fluctuation at all peaks. We performed a series of affine transformations on the energy-filtered vibrational signal maps to correct the sample drift for each u.c. based on the atomic structure in independently acquired aHAADF images[40] (see more details in Methods and Extended Data Figs. 5b and 5c). The u.c.-averaged signal mapping results from 10 meV to 110 meV with an energy bin width of 10 meV in Fig. 2e all show clear atomic resolution features and are in good agreement with simulated energy-filtered vibrational mapping results using a modified frequency-resolved frozen phonon multislice method[41,42] in Fig. 2f (see more details in Methods).

As marked in Fig. 2b, Sr atoms reside at the corner of the STO unit cell, a mixture of Ti and O1 atoms are in the middle, and the O2 and O3 atoms reside at the middle of four sides. Thus, Sr, Ti/O1, O2, and O3 atomic columns are spatially separated in the [001]-projected structure. We also differentiate between three oxygen atoms (O1, O2, and O3 in Fig. 2b) to emphasize their site-specific difference. When shifting along X' direction, the 10–20 meV mapping presents a strong vibrational intensity at the Sr position, matching with the simulated signal map at 12.4 meV (3 THz) and the Sr-projected local PhDOS. In medium energy ranges (20-30 meV, 30-40meV, 40-50 meV, and 50-60 meV), the position of brighter atoms changes from Sr to Ti/O1 columns, representing a mixed vibrations of Ti and O atoms. Moreover, as the mapping energy range increases to higher values, the surrounding oxygen atoms (O2 and O3) are gradually brighter and form four tails around the central Ti/O1 columns, which are more pronounced in the raw simulation results (Extended Data Fig. 6). By contrast, in even higher energy ranges (60-70 meV, 90-100 meV, and 100-110 meV), some of the surrounding oxygen atoms show the strongest intensities, while the middle Ti/O1 columns and Sr atoms are all dimmer. These phenomena can be explained by oxygen-dominated vibrational modes from 60 to 110 meV according to the PhDOS. The vibrational signal mappings for shifting along Y' direction in Fig. 2i exhibit similar atomic patterns, also matching with the simulation results in Fig. 2j. For both dark-field conditions, this vibrational EELS mapping method reveals the atomic features and provides the elemental-specific mapping in different energy ranges in accordance with varied contributions from each atom in PhDOS.

More fundamentally, this observation reflects the atomic-level quantum mechanical pictures of phonon as a wavefunction. Phonons are believed to exist uniformly in perfect crystalline materials at equilibrium conditions. Imperfections can generate localized phonon modes near the defective regions as a perturbation in the bulk phonon structure according to theories and plentiful recent STEM-EELS experiments[13–22]. However, the origin of local variation of vibrational EELS signals at individual atom level in a perfect lattice is still doubted. The atomic resolution features in BN[23] and Si[24] can be interpreted as the thickness effect that electrons will experience more inelastic scattering at atomic columns and yield higher phonon signals but less at the position between atomic columns. Our results clearly prove that the vibrational EELS signals have varied intensities at different atomic columns including pure Sr, mixed Ti/O, and pure O columns at different energy ranges, indicating the variation of corresponding wavefunctions (or phonon polarization vectors) at distinct atomic columns. Therefore, the cartographic delineation of phonon eigenvectors using our method offers direct evidence of the existence of atomic-level fluctuations for phonon wavefunctions.

Beyond the atomic features, there are notable differences between X'- and Y'-shifted dark-field EELS results. For the high energy oxygen-related phonon states at $D_{EEA}$ = (-62, 0), only equatorial oxygen (O3) atoms possess pronounced vibrational signals, while O1 and O2 have low vibrational signals. In contrast, at $D_{EEA}$ = (0, 62), only apical oxygen (O2) atoms have brighter contrast, while O1 and O3 atoms are silent. Thus, with orthogonal displacement vectors, the acquired vibrational signals for oxygen atoms in STO exhibit entirely different intensities. Looking deeper, we also investigated the vibrational intensities of all oxygen atoms in other energy ranges. Figure 3a compares the experimental and simulated line profiles for O2 and O3 atoms in two representative energy ranges. In the 40-50 meV energy range, the surrounding oxygen atoms near the central Ti/O1 column possess varied intensities. O2 atoms are brighter for X' shift, while O3 atoms are brighter for Y' shift, in line with the relative in-plane motion of Ti and O atoms of TO3/LO2 modes. The line profiles in simulation results show perfect consistency with experimental intensity modulations (Fig. 3a). In the 100-110 meV energy range, the line profiles express an opposite tendency for the intensity modulations between O2 and O3 atoms. O3 atoms become brighter for X' shift, while O2 atoms become brighter for Y' shift. This is roughly understandable as high-energy LO4 phonons manifest as the breathing modes of oxygen octahedra[43], with O1, O2, and O3 oscillating along Z, Y, and X axes, respectively.

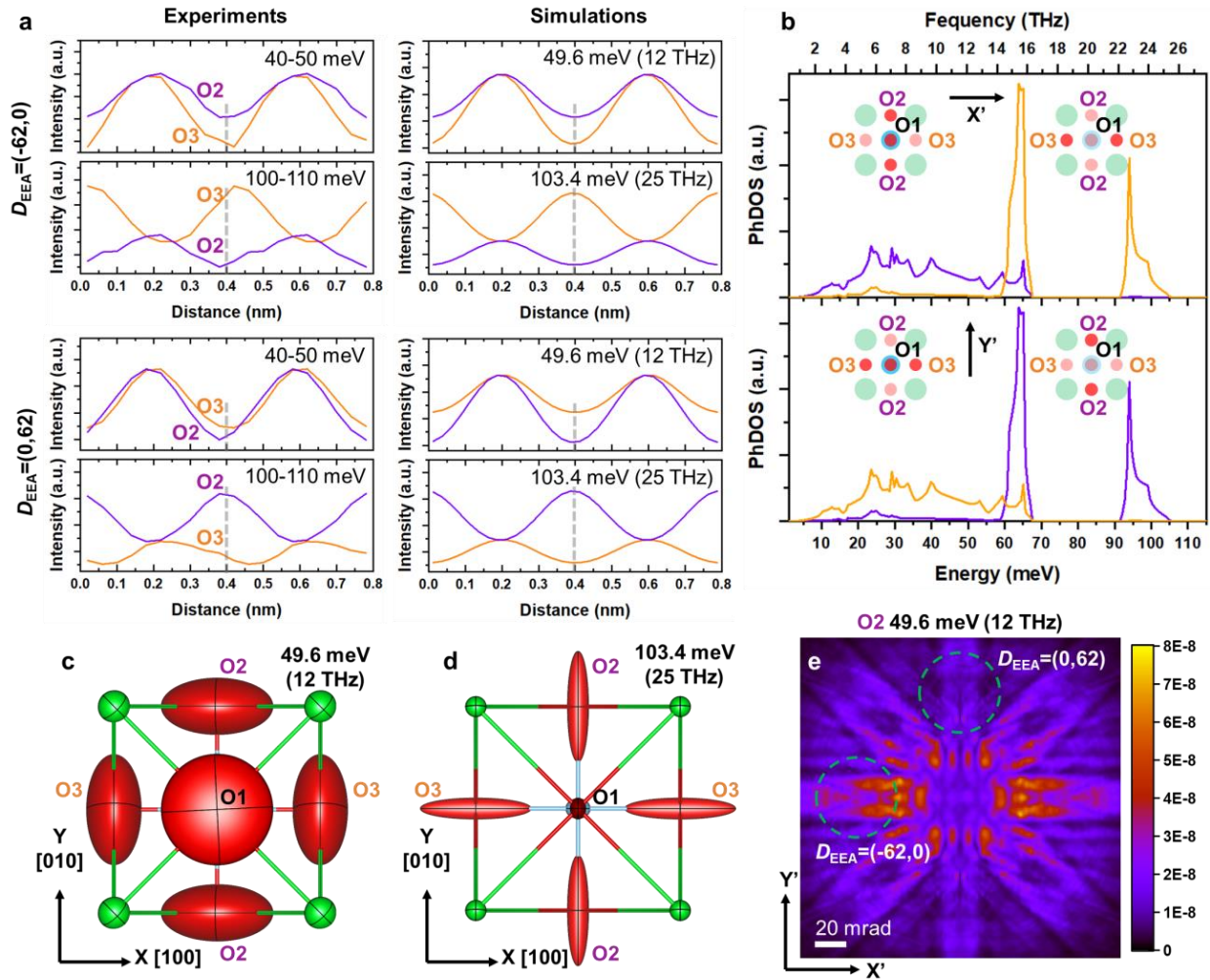

**Fig. 3| Selective excitation and frequency-dependent anisotropy of oxygen vibrations. a,** Line profiles of vibrational signals in chosen energy ranges along horizontal (orange) and vertical (purple) directions as indicated in Fig. 2. O2 and O3 atoms are located at 0.4 nm position as indicated by gray dashed lines. **b,** Projected PhDOS of O1, O2, and O3 atoms along X' (upper) and Y' (lower) directions. The O1 curve overlaps with the O2 (O3) one in the upper (lower) panel. Insets illustrate the selective excitation of different oxygen atoms in different energy ranges. **c,d,** [001]-projected thermal ellipsoids representing anisotropy of individual mean-squared displacements of all atoms at 49.6 meV (12 THz, **c**) and 103.4 meV (25 THz, **d**), respectively. The size of displacement is exaggerated for visual clarity. **e,** An energy filtered diffraction pattern of inelastically scattered electrons with an energy loss of 49.6 meV (12 THz) when the electron beam passes through at an O2 column. Two dashed green circles indicate the EEA position. Scale bar is 20 mrad.

Such contrasting intensity variations can be theoretically explained by considering the scattering probability of fast electrons by lattice vibrations and the atom projected PhDOS along different directions. The selection rule of vibrational EELS signal stems from the term $\boldsymbol{q}\cdot\boldsymbol{e}_i(\omega)$, where $\boldsymbol{q}$ is the momentum exchange of scattered electron ($\sim\boldsymbol{D}_{EEA}$) and $\boldsymbol{e}_i(\omega)$ is the phonon eigenvector of atom *i* at a frequency $\omega$ [19,44]. Since vibrational EELS is primarily sensitive to the phonon modes vibrating in the plane perpendicular to the electron beam[10, 44], we only consider the in-plane components $q_x$ and $q_y$ along X' and Y' directions and plot the local PhDOS projected along $e_x$ and $e_y$ in Fig. 3b. For the X' component, O1 and O2 atoms possess almost the same PhDOS, but O3 is totally different. Below 60 meV (14.5 THz), the PhDOS of O1 and O2 atoms is higher than O3 atoms, leading to the brighter contrast at O2 and Ti/O1 atomic columns in the vibrational signal mapping results. Above 60 meV, the PhDOS of O3 atoms is higher and even dominant in the 90-110 meV energy range, leading to the brighter contrast at O3 atoms and silent O2 and Ti/O1 atomic columns in the mapping results. Conversely, the oxygen projected PhDOS along Y' also explains the observed changes in corresponding mapping results. Therefore, the contrast variation comes from the anisotropy of phonon eigen-displacement of different oxygen atoms. We further quantify the intensity difference between O3 and O2 atoms for all energy ranges and compare them with simulated values (Extended Data Fig. 7). Simulation results point out that for X' shift, the O3 intensity is lower than that of O2 at energies lower than 60 meV, and the O3 intensity is higher than O2 at energies higher than 60 meV. The intensity contrast between O3 and O2 reaches the lowest value of -0.13 at 49.6 meV, and the highest value of 0.34 at 103.4 meV. For Y' shift, the contrast trend flips. The experimental contrast follows a similar relationship (Extended Data Fig. 7b), and their absolute values reach the maximums of 0.67 and 0.35 for X' and Y' shifts, respectively, in the 100–110 meV mapping results.

The anisotropy of atomic vibrations can be visualized by plotting the mean-square displacements of individual atoms, also referred to as thermal ellipsoids. Figure 3c depicts obvious vibrational anisotropies at both O2 and O3 atoms as evidenced by their ellipsoidal shapes. At 49.6 meV, the long axis of O2 (O3) ellipsoids is parallel to the X (Y) axis, and thus the corresponding vibrational signal is stronger when coupling with electron beam deflected along X' (Y'). At 103.4 meV, the thermal ellipsoids of both O2 and O3 rotate by 90°, giving rise to different contrasts in mapping results. Moreover, the elongations of O2 and O3 ellipsoids are larger at 103.4 meV, which is consistent with the strongest contrast difference in the 100-110 meV energy range. We show more plots of thermal ellipsoids for an extended selection of phonon energies as well as the total PhDOS (Extended Data Fig. 8). All Sr and Ti atoms exhibit spherical shapes indicating the homogeneity of corresponding phonon eigenvectors, matching with the similar intensity at Sr and Ti/O1 atomic columns between X' and Y' shifts in experiments. From the viewpoint of symmetry, STO has a centrosymmetric cubic structure (*Pm-3m*) at room temperature, and one does not expect any macroscopic anisotropic properties[5]. However, only Sr and Ti sites obey the full point group symmetry ($O_h$), while the oxygen atoms possess a lowered site-symmetry ($D_{4h}$). Thus, the O1 site exhibits a four-fold rotation symmetry ($C_4$) along [001] direction, while the O2 and O3 sites are only invariant under mirroring along [001]. The lowered point group symmetry at O2 atoms is also reflected in the simulated energy filtered diffraction pattern in Fig. 3e. The asymmetric intensity distribution along X' and Y' directions will give rise to the difference in the dark-field vibrational EELS signals from two different deflected EEA. Similar asymmetric energy filtered diffraction patterns are also observed at O3 columns, while Sr and Ti columns yield a symmetric energy filtered diffraction pattern with four-fold rotation symmetry (Extended Data Fig. 9). Although X-ray diffraction and many other methods have been employed to estimate the anisotropy of oxygen atoms in STO, the obtained thermal ellipsoids are averaged out over the entire energy span[5,8,11]. These methods cannot separate the varied thermal ellipsoids at different frequencies, which are crucial for understanding the thermal and photonic responses related to acoustic and optical branches, respectively. The thermal ellipsoids of oxygen atoms have an obvious change from low to high energy phonon modes (Extended Data Fig. 8). The oblate ellipsoids at low phonon energies (3–12 THz) including the soft-phonon modes indicate the oxygen atoms are vibrating in the Sr-O plane and provide a major contribution to the heat transport. This behavior is also prone to the libration of TiO$_6$ octahedra, which is a precursor of the antiferrodistortive phase transition below 105 K[8,11]. Further study of the energy shift of the lowest phonon modes may reveal the atomic origin of ferroelectricity according to the soft mode theory[33,34]. The high-energy LO4 phonon eigenvectors exhibit a distinct prolate ellipsoids, showing that oxygen atoms are vibrating along Ti-O bonding direction, and can be coupled with mid-infrared optical lasers to generate a photoflexoelectic response[28].

In conclusion, the reported dark-field vibrational EELS approach with a momentum-selective displacement vector of EELS entrance aperture achieves an element-resolved and atomic site-specific detection of phonon eigen-displacements in a complex perovskite STO single crystal. This method further reveals frequency-dependent vibrational anisotropies along different directions. We found that oxygen atoms in STO possess anisotropic vibrations along [100] or [010] directions at medium and high phonon energies, depending on the locally lowered point group symmetry, while all Sr and Ti atoms exhibit isotropic atomic vibrations. These results pave a new avenue to investigating the phonon anisotropy with unprecedented atomic resolution for a variety of general dielectric materials to understand their optical, thermal, and thermoelectric properties. The selectivity can be modified by rotating the displacement vector of the EEA ($D_{EEA}$). For example, <111> direction should be used for observing spontaneous polarization of BiFeO$_3$ in a pseudo-cubic phase[45]. Such features can be correlated with the ferroelectric phase transition of perovskite oxides and the origin of ferroelectricity at varied temperatures. By sampling more displacement vectors in the reciprocal space, vibrational EELS method can be upgraded to a general spectroscopic ellipsometry.

1   Schubert, M. *et al.* Anisotropy, phonon modes, and free charge carrier parameters in monoclinic β-gallium oxide single crystals. *Phys. Rev. B* **93**, 125209 (2016).
2   Ma, W. L. *et al.* In-plane anisotropic and ultra-low-loss polaritons in a natural van der Waals crystal. *Nature* **562**, 557-562 (2018).
3   Hu, G. W. *et al.* Topological polaritons and photonic magic angles in twisted α-MoO$_3$ bilayers. *Nature* **582**, 209-213 (2020).
4   Kim, S. E. *et al.* Extremely anisotropic van der Waals thermal conductors. *Nature* **597**, 660-665 (2021).
5   Lin, I. C. *et al.* Extraction of Anisotropic Thermal Vibration Factors for Oxygen from the Ti $L_{2,3}$-Edge in SrTiO$_3$. *The Journal of Physical Chemistry C* **127**, 17802-17808 (2023).
6   Bubnova, R., Volkov, S., Albert, B. & Filatov, S. Borates—Crystal Structures of Prospective Nonlinear Optical Materials: High Anisotropy of the Thermal Expansion Caused by Anharmonic Atomic Vibrations. *Crystals* **7**, 93 (2017).
7   Wenzel, L., Stöhr, J., Arvanitis, D. & Baberschke, K. Vibrational anisotropy and anharmonicity of N atoms bonded to Ni(100). *Phys. Rev. Lett.* **60**, 2327-2330 (1988).
8   Abramov, Y. A., Tsirelson, V. G., Zavodnik, V. E., Ivanov, S. A. & Brown, I. D. The chemical bond and atomic displacements in SrTiO$_3$ from X-ray diffraction analysis. *Acta Cryst. B* **51**, 942-951 (1995).

9   Gong, Y. *et al.* Polarized Raman Scattering of In-Plane Anisotropic Phonon Modes in α-MoO$_3$. *Adv. Opt. Mater.* **10**, 2200038 (2022).
10  Yan, X., Gadre, C. A., Aoki, T. & Pan, X. Probing molecular vibrations by monochromated electron microscopy. *Trends Chem.* **4**, 76-90 (2022).
11  Jauch, W. & Reehuis, M. Electron-density distribution in cubic SrTiO$_3$: a comparative gamma-ray diffraction study. *Acta Cryst. A* **61**, 411-417 (2005).
12  Krivanek, O. L. *et al.* Vibrational spectroscopy in the electron microscope. *Nature* **514**, 209-212 (2014).
13  Hage, F. S., Radtke, G., Kepaptsoglou, D. M., Lazzeri, M. & Ramasse, Q. M. Single-atom vibrational spectroscopy in the scanning transmission electron microscope. *Science* **367**, 1124-1127 (2020).
14  Xu, M. *et al.* Single-atom vibrational spectroscopy with chemical-bonding sensitivity. *Nat. Mater.* **22**, 612-618 (2023).
15  Yan, X. *et al.* Single-defect phonons imaged by electron microscopy. *Nature* **589**, 65-69 (2021).
16  Cheng, Z. *et al.* Experimental observation of localized interfacial phonon modes. *Nat. Commun.* **12**, 6901 (2021).
17  Qi, R. *et al.* Measuring phonon dispersion at an interface. *Nature* **599**, 399-403 (2021).
18  Hoglund, E. R. *et al.* Emergent interface vibrational structure of oxide superlattices. *Nature* **601**, 556-561 (2022).
19  Gadre, C. A. *et al.* Nanoscale imaging of phonon dynamics by electron microscopy. *Nature* **606**, 292-297 (2022).
20  Zeiger, P. M. & Rusz, J. Simulations of spatially and angle-resolved vibrational electron energy loss spectroscopy for a system with a planar defect. *Phys. Rev. B* **104**, 094103 (2021).
21  Hoglund, E. R. *et al.* Direct Visualization of Localized Vibrations at Complex Grain Boundaries. *Adv. Mater.* **35**, e2208920 (2023).
22  Haas, B. *et al.* Atomic-Resolution Mapping of Localized Phonon Modes at Grain Boundaries. *Nano Lett.* **23**, 5975-5980 (2023).
23  Hage, F. S., Kepaptsoglou, D. M., Ramasse, Q. M. & Allen, L. J. Phonon Spectroscopy at Atomic Resolution. *Phys. Rev. Lett.* **122**, 016103 (2019).
24  Venkatraman, K., Levin, B. D. A., March, K., Rez, P. & Crozier, P. A. Vibrational spectroscopy at atomic resolution with electron impact scattering. *Nat. Phys.* **15**, 1237-1241 (2019).
25  Sirenko, A. A. *et al.* Soft-mode hardening in SrTiO$_3$ thin films. *Nature* **404**, 373-376 (2000).
26  Stengel, M. & Spaldin, N. A. Origin of the dielectric dead layer in nanoscale capacitors. *Nature* **443**, 679-682 (2006).
27  Huang, J. K. *et al.* High-kappa perovskite membranes as insulators for two-dimensional transistors. *Nature* **605**, 262-267 (2022).
28  Nova, T. F., Disa, A. S., Fechner, M. & Cavalleri, A. Metastable ferroelectricity in optically strained SrTiO$_3$. *Science* **364**, 1075-1079 (2019).
29  Hong, C. *et al.* Anomalous intense coherent secondary photoemission from a perovskite oxide. *Nature* **617**, 493-498 (2023).
30  Muller, D. A. *et al.* Atomic-scale chemical imaging of composition and bonding by aberration-corrected microscopy. *Science* **319**, 1073-1076 (2008).
31  Gao, W. P. *et al.* Real-space charge-density imaging with sub-ångström resolution by four-dimensional electron microscopy. *Nature* **575**, 480-484 (2019).
32  Ishikawa, R., Tanaka, R., Kawahara, K., Shibata, N. & Ikuhara, Y. Atomic-Resolution Topographic Imaging of Crystal Surfaces. *ACS Nano* **15**, 9186-9193 (2021).

33  Scott, J. F. Soft-Mode Spectroscopy - Experimental Studies of Structural Phase-Transitions. *Rev. Mod. Phys.* **46**, 83-128 (1974).
34  Casella, L. & Zaccone, A. Soft mode theory of ferroelectric phase transitions in the low-temperature phase. *J. Phys. Condens. Matter* **33**, 165401 (2021).
35  Lee, J. J. *et al.* Interfacial mode coupling as the origin of the enhancement of $T_c$ in FeSe films on $SrTiO_3$. *Nature* **515**, 245-248 (2014).
36  Ji, D. X. *et al.* Freestanding crystalline oxide perovskites down to the monolayer limit. *Nature* **570**, 87-90 (2019).
37  He, R. *et al.* Structural phase transitions in $SrTiO_3$ from deep potential molecular dynamics. *Phys. Rev. B* **105**, 064104 (2022).
38  van der Marel, D., Barantani, F. & Rischau, C. W. Possible mechanism for superconductivity in doped $SrTiO_3$. *Phys. Rev. Res.* **1**, 013003 (2019).
39  Niedermeier, C. A. *et al.* Phonon scattering limited mobility in the representative cubic perovskite semiconductors $SrGeO_3$, $BaSnO_3$, and $SrTiO_3$. *Phys. Rev. B* **101**, 125206 (2020).
40  Smith, J., Huang, Z., Gao, W., Zhang, G. & Chi, M. Atomic Resolution Cryogenic 4D-STEM Imaging via Robust Distortion Correction. *ACS Nano* **17**, 11327-11334 (2023).
41  Zeiger, P. M. & Rusz, J. Efficient and Versatile Model for Vibrational STEM-EELS. *Phys. Rev. Lett.* **124**, 025501 (2020).
42  Zeiger, P. M. & Rusz, J. Frequency-resolved frozen phonon multislice method and its application to vibrational electron energy loss spectroscopy using parallel illumination. *Phys. Rev. B* **104**, 104301 (2021).
43  Cancellieri, C. *et al.* Polaronic metal state at the $LaAlO_3$/$SrTiO_3$ interface. *Nat. Commun.* **7**, 10386 (2016).
44  Nicholls, R. J. *et al.* Theory of momentum-resolved phonon spectroscopy in the electron microscope. *Phys. Rev. B* **99**, 094105 (2019).
45  Nelson, C. T. *et al.* Domain Dynamics During Ferroelectric Switching. *Science* **334**, 968-971 (2011).

**Methods**

**Sample Preparation.** A water-soluble $Sr_3Al_2O_6$ (SAO) film was first grown on a (001) $TiO_2$-terminated STO single-crystalline substrate followed by the growth of a thin STO film by oxide molecular beam epitaxy. During the growth, reflection high energy electron diffraction (RHEED) was employed to monitor the surface quality and to precisely control the thickness of the films according to RHEED intensity oscillations. A 6 u.c.-thick SAO layer was first deposited at a substrate temperature of 850 °C with an oxygen pressure of $1\times10^{-6}$ Torr. A 50 u.c.-thick STO layer was subsequently deposited at a relatively lower temperature of 650 °C to reduce the diffusion of Ti atoms into SAO layers. To prepare TEM specimens, the STO/SAO/STO film was immersed into deionized water at room temperature until the sacrificial SAO layer was completely dissolved, and the peeled STO film was floating on the water surface. Then, the STO film was picked up by lacey carbon TEM grids covered with a few-layer graphene film under an optical microscope. The STO/Graphene TEM specimen was baked for over 24 hours at 160 °C before inserting into a microscope.

**Experimental Details.** Low magnification TEM images and selected area diffraction patterns were obtained using a JEOL 2100F TEM operating at 200 kV. All STEM images and dark-field vibrational EELS spectra were collected by Nion UltraSTEM 200 equipped with C3/C5 aberration corrector and high-energy resolution monochromated EELS system (HERMES). The convergence semi-angle of the 60 kV electron probe was 33 mrad with a beam current of ~100 pA before monochromation. The inner and outer collection semi-angles of HAADF detector were 95 mrad and 210 mrad, respectively. The dwell time of HAADF images without monochromation is 8 μs, while that with monochromation is 20 μs. The collection semi-angle of EEA was 25 mrad. The microscope condition was carefully tuned to center electron probe, HAADF detector, and EEA with Ronchigram camera before changing to off-axis EELS settings. [100] and [010] directions of observed STO films were identified from atomic resolution STEM images and used to set up the displacement direction of the diffraction pattern on the Ronchigram camera via post-specimen projector lenses. To accurately correlate the directions in the diffraction plane with crystalline orientations, we considered and measured the rotation between image and diffraction pattern by comparing the same sample feature (e.g., edges) in an underfocused Ronchigram image with that in a low magnification focused STEM image[46]. The displacement distance ($D_{EEA}$) was set to 62 mrad to avoid any overlapping between center disk and EEA or between center disk and HAADF detector. An alpha-type monochromator was activated to obtain the best energy resolution of 5.7 meV under the regular on-axis condition. Monochromator slit aberrations, monochromator exit aberrations, probe aberrations, projector aberrations, and EELS aberrations were all corrected to guarantee sufficient spatial resolution and the best and stable energy resolution after monochromation. Most importantly, to routinely achieve a good energy resolution (9–11 meV), which is measured as the full width at half maximum (FWHM) of the zero-loss peak (ZLP), under dark-field EELS condition, the second order EELS aberrations were tuned to be less than 100 V, and the third order EELS aberrations were tuned to be less than 1000 V. The EELS dispersion was set to about 0.5 meV/channel and frequently calibrated by wobbling the drift tube.

     Atomic resolution hyperspectral images of vibrational signals were acquired by running a built-in "Spectrum Imaging" function in the Nion's Swift software. The field of view of mapping area was 1.5–2.0 nm containing 4–5 u.c. of STO with 45–60 pixels in each direction. Each spectrum was acquired as a single frame with 1 s exposure time. The total acquisition time of

hyperspectral image dataset was 30–60 min to ensure a decent probe current without a significant decay and to reduce the influence of sample drifting. The datasets used for extracting atomic resolution images must have a sample drift less than 0.8 nm (~2 u.c. of STO) without noticeable contamination. One example is shown in Extended Data Fig. 4.

**Data Analysis and Drift Correction.** Hyperspectral image datasets were aligned by the ZLP center and normalized by the ZLP height. To improve the signal-to-noise ratio and make the data analysis more reliable, all hyperspectral image datasets were binned by every 4×4 pixel area. Then, background subtraction was carefully conducted by setting three fitting windows and fitting a power law function as $I(E) = aE^{-b} + c$, where $I(E)$ is the intensity of vibrational spectrum, $E$ is the energy loss in unit of meV, and $a$, $b$, and $c$ are fitting coefficients[47]. The fitting windows for most STO results are 7–9, 75–85, and 110–130 meV to match with the energy span of major peaks in calculated PhDOS and energy band gap (70–90 meV).

Although we tried to reduce the sample drifting and charging issues, there are still some inevitable lattice distortions in most vibrational EELS mapping datasets as shown in Extended Data Fig. 4. To correct such distortions, a standard affine transformation algorithm was utilized on each unit cell. The drifting compensation was achieved by reconstructing the unit cell using the estimated affine transformation matrix obtained from the OpenCV package[48], where the random sample consensus method is implemented such that:

$$\begin{bmatrix} x \\ y \end{bmatrix} = \begin{bmatrix} a_{11} & a_{12} \\ a_{21} & a_{22} \end{bmatrix} \begin{bmatrix} X \\ Y \end{bmatrix} + \begin{bmatrix} b_1 \\ b_2 \end{bmatrix},$$

where $x$ and $y$ are the coordinates of the reference position (Sr atoms in our case), $X$ and $Y$ are the actual atom position, and the transformation matrix $H$ is defined as $H = \begin{bmatrix} a_{11} & a_{12} & b_1 \\ a_{21} & a_{22} & b_2 \end{bmatrix}$. First, a 2D Gaussian fitting method in the Atomap software[49] was employed to identify the position of Sr atoms in both aHAADF-STEM images and the energy-filtered phonon map of 10–20 meV, serving as reference positions and measured positions respectively. By combining prediction based on the Ti atom positions in the 20–30 meV maps and manual selection, a series of discernible unit cells were marked with intact Sr atoms at the four corners of unit cells in the 10–20 meV maps and visible Ti atoms in the center of the unit cells in the 20–30 meV map. Then, a series of transformation matrixes $H$ were calculated by comparing measured positions of four Sr atoms in selected unit cells with their reference positions. All points inside individual unit cells in the original image were mapped to their corresponding positions in the corrected images using the transformation matrix $H$ obtained on this unit cell. The same $H$ was also applied to correct other energy-filtered vibrational signal maps in the same hyperspectral image dataset. To address the sampling difference between the corrected and original images, bilinear interpolation was applied to individual vibrational signal maps. This interpolation method estimates the pixel values at non-integer positions by considering the intensities of neighboring pixels, ensuring a smooth and accurate representation of the original image during the transformation process. The resulting unit cells were then remapped back to their original undistorted cubic shape as shown in Extended Data Figs. 5b and 5c.

**First-Principles Calculations.** The phonon dispersion and density of states simulations of STO at 0 K were first calculated using the Vienna ab-initio simulation package (VASP) at the level of the spin-polarized generalized-gradient approximation (GGA) with the functional developed by Perdew-Burke-Ernzerhof[50]. The framework of the projector augmented wave (PAW) method was adopted to treat the interaction between valence electrons and ionic cores[51,52]. In all the density functional theory (DFT) calculations, the energy cutoff for the plane wave basis expansion was set to 700 eV, and the criterion for total energy convergence is set to $10^{-6}$ eV. A 4×4×4 supercell (320 atoms/supercell) was used to mimic STO, and all atoms were fully relaxed using the conjugated gradient method for the energy minimization until the force on each atom became smaller than 0.01 eV/Å. The phonon dispersion and density of states were further obtained using density-functional perturbation theory (DFPT)[53].

**Phonon Dispersion Calculations and Molecular Dynamics Simulations Using DeepMD.**
Molecular dynamics (MD) simulations were carried out using LAMMPS with the DeepMD potential for cubic STO[37,54]. The potential successfully describes a structural phase transition from the low temperature tetragonal phase to the high temperature cubic phase. The phonon dispersion and PhDOS of the DeepMD potential at 0 K and 300 K were calculated with the static displacement method using the phonoLAMMPS, DynaPhoPy, and Phonopy packages[55–58]. The supercell was thereby set to 2×2×2 u.c.. The k-mesh for the integration of the PDOS was set to 77×77×77. The crystal's microscopic anharmonic properties were taken into account using the DynaPhoPy software, integrated into the phonoLAMMPS interface. To accurately predict the LO-TO splitting at the Γ point, a non-analytic correction was applied to the final phonon dispersion and PhDOS calculations. The interpolation scheme in the Phonopy package utilized the dielectric constant and born effective charge obtained from the DFT calculation. This comprehensive approach ensures precise calculations of the phonon dispersion curve and density of states, incorporating both the anharmonic properties and non-analytic corrections.

To simulate vibrational EELS, a series of MD simulations were performed at 300 K, when STO is in its cubic phase. The supercell for the MD calculations consisted of 12×12×60 u.c.. The supercell dimensions of 46.906×46.906×234.528 Å$^3$ were computed from the time-average of the fluctuating supercell dimensions in a constant temperature and constant pressure MD simulation (NPT ensemble) at a temperature of 300 K and a pressure of 0.0 bar. A time step of 1 fs was used in all MD runs. After the box size had been fixed, a Langevin thermostat with a damping parameter of 100 fs was used to equilibrate the temperature to 300 K in a constant temperature and constant volume MD simulation (NVT ensemble). This simulation was carried out in order to sample a set of 100 representative states of the vibrating atomic system, consisting of atomic positions and atomic velocities. These representative states of the system were extracted every 2000 time steps (2 ps) after an initial equilibration of 5000 time steps (5 ps), guaranteeing that they are uncorrelated.

The 100 representative states of the atomic system in the NVT ensemble served as initial inputs to compute 100 trajectories in a constant energy MD simulation (micro-canonical ensemble, NVE ensemble). To form each NVE trajectory we ran the simulation for 5000 time steps (5 ps) and sampled then the positions of all atoms every 10 time steps (10 fs) to obtain 5000 position samples of all atoms. We further split each of the NVE trajectories into two parts of 2500 position samples each, and the resulting 200 trajectories were used for the simulation of vibrational EELS in the following section.

**Frequency-Resolved Frozen Phonon Multislice Simulations.** The electron scattering simulations in Figs. 2 and 3, and Extended Data Fig. 6 are performed using the frequency-resolved frozen phonon multislice (FRFPMS) method, in which the inelastic signal is computed as the difference between the incoherent and the coherent averages of beam exit wave functions computed over structure snapshots, in which only displacements due to vibrational modes within a narrow range of frequencies are present[20,41,42].

In order to extract structure snapshots representing certain phonon excitations for the cubic phase of STO, we implemented band-pass filtering on the 200 NVE trajectories consisting of 2500 position samples. That is, for a selected NVE trajectory we obtained the Fourier components of the (time-dependent) position samples of all atoms using a discrete Fourier Transform, then set all Fourier components to zero except for those inside a desired, narrow range of frequencies, and finally performed an inverse Fourier Transform to bring back the band-pass filtered atomic position samples into the time domain. In this work we considered frequency bands centered at frequencies from 1 to 25 THz in a step of 1 THz. Each band covered a frequency range of ±0.5 THz around said center frequency. This is sufficient to encompass the whole range of vibrational frequencies (larger than 0.5 THz) where the PhDOS of STO is non-zero. From each of the 200 band-pass filtered trajectories we have extracted one structure snapshot for each of the 25 frequencies.

These structure snapshots were used in the FRFPMS method to calculate the inelastically scattered intensity in each frequency bin and to determine the thermal displacement ellipsoids shown in Figs. 3c and 3d, and Extended Data Fig. 8. All multislice calculations required for the FRFPMS method were carried out using the DrProbe software package[59]. The settings for the accelerating voltage, convergence semi-angle, EEA size, and defocus in the multislice calculations closely followed the experimental conditions. The lateral grid size for the calculations was 576×576 and the supercell has been divided into 600 approximately 0.4-Å thick slices. In the multislice runs we have assumed an isotropic Debye-Waller factor and absorptive potential, both parametrized by Biso values of 0.5448, 0.2948, and 0.6711 Å$^2$ for Sr, Ti, and O atoms, respectively. These values were determined from the mean squared displacements (MSD) in the MD trajectories.

46  Spiecker, E. Determination of crystal polarity from bend contours in transmission electron microscope images. *Ultramicroscopy* **92**, 111-132 (2002).
47  Yan, X. *et al.* Curvature-Induced One-Dimensional Phonon Polaritons at Edges of Folded Boron Nitride Sheets. *Nano Lett.* **22**, 9319-9326 (2022).
48  Culjak, I., Abram, D., Pribanic, T., Dzapo, H. & Cifrek, M. in *2012 proceedings of the 35th international convention MIPRO.*  1725-1730 (IEEE).
49  Nord, M., Vullum, P. E., MacLaren, I., Tybell, T. & Holmestad, R. Atomap: a new software tool for the automated analysis of atomic resolution images using two-dimensional Gaussian fitting. *Advanced Structural and Chemical Imaging* **3**, 9 (2017).
50  Perdew, J. P., Burke, K. & Ernzerhof, M. Generalized gradient approximation made simple. *Phys. Rev. Lett.* **77**, 3865-3868 (1996).
51  Blöchl, P. E. Projector Augmented-Wave Method. *Phys. Rev. B* **50**, 17953-17979 (1994).
52  Kresse, G. & Joubert, D. From ultrasoft pseudopotentials to the projector augmented-wave method. *Phys. Rev. B* **59**, 1758-1775 (1999).
53  Gonze, X. & Lee, C. Dynamical matrices, born effective charges, dielectric permittivity tensors, and interatomic force constants from density-functional perturbation theory. *Phys. Rev. B* **55**, 10355-10368 (1997).
54  Thompson, A. P. *et al.* LAMMPS - a flexible simulation tool for particle-based materials modeling at the atomic, meso, and continuum scales. *Comput. Phys. Commun.* **271**, 108171 (2022).
55 Carreras, A. *phonoLAMMPS Documentation*, <https://github.com/abelcarreras/phonolammps> (2023).

56  Carreras, A., Togo, A. & Tanaka, I. DynaPhoPy: A code for extracting phonon quasiparticles from molecular dynamics simulations. *Comput. Phys. Commun.* **221**, 221-234 (2017).
57  Togo, A., Chaput, L., Tadano, T. & Tanaka, I. Implementation strategies in phonopy and phono3py. *J. Phys. Condens. Matter* **35**, 353001 (2023).
58  Togo, A. First-principles Phonon Calculations with Phonopy and Phono3py. *J. Phys. Soc. Jpn.* **92**, 012001 (2023).
59  Barthel, J. Dr. Probe: A software for high-resolution STEM image simulation. *Ultramicroscopy* **193**, 1-11 (2018).
60  Momma, K. & Izumi, F. VESTA 3 for three-dimensional visualization of crystal, volumetric and morphology data. *J. Appl. Crystallogr.* **44**, 1272-1276 (2011).

**Data availability**

The datasets generated during and/or analysed during the current study are available from the corresponding authors on reasonable request.

**Acknowledgments** The experimental work was supported by the Department of Energy (DOE), Office of Basic Energy Sciences, Division of Materials Sciences and Engineering under Grant DE-SC0014430. The authors acknowledge the use of facilities and instrumentation at the UC Irvine Materials Research Institute (IMRI), which is supported in part by the National Science Foundation through the UC Irvine Materials Research Science and Engineering Center (DMR-2011967). J.R. and P.Z. acknowledge the Swedish Research Council, Olle Engkvist's foundation, Carl Trygger's Foundation, Knut and Alice Wallenberg Foundation for financial support. Simulations were enabled by resources provided by the National Academic Infrastructure for Supercomputing in Sweden (NAISS) and the Swedish National Infrastructure for Computing (SNIC) at NSC Centre partially funded by the Swedish Research Council through grant agreements no. 2022-06725 and no. 2018-05973. We thank Dr. Mingjie Xu and Dr. Sung Joo Kim (University of California, Irvine, CA, USA) for the wet transfer of freestanding films, thank Dr. Tracy C. Lovejoy and Dr. Niklas Dellby (Nion R&D, Kirkland, WA, USA) for microscope alignment and high order EELS aberration correction, and thank Prof. Bolin Liao (University of California, Santa Barbara, CA, USA) for valuable suggestions.

**Author contributions** X.P. conceived this project and designed the studies, with further contributions from J.R. and X.Y.. X.Y. performed STEM-EELS experiments and analyzed all datasets with the help of Y.H., H.Y., C.A.G., and T.A.. Y.H. performed drift correction. P.Z. and J.R. designed and performed the molecular dynamic-based calculation of vibrational spectra with the help of Y.H., R.H., and Z.Z.. J.L. and R.W. conducted first-principles calculations. The freestanding STO films were provided by H.S. and Y.N. All authors discussed and commented on the results. The manuscript was prepared by X.Y., P.Z., J.R., and X.P., with contributions from all other co-authors.

**Competing interests:** The authors declare no competing interests.